\documentclass[12pt]{article}
\usepackage{graphicx}

\begin{document}


\title{\bf
Monopole Solutions of the Massive SU(2) Yang-Mills-Higgs Theory\footnote{to be submitted for publication}}

\author{
{\bf Rosy Teh\footnote{e-mail: rosyteh@usm.my}, Khai-Ming Wong, and Pin-Wai Koh}\\
{\normalsize School of Physics, Universiti Sains Malaysia}\\
{\normalsize 11800 USM Penang, Malaysia}}

\date{February, 2010}
\maketitle

\begin{abstract} 
Monopoles in topologically massive gauge theories in 2+1 dimensions with a Chern-Simon mass term have been studied by Pisarski some years ago. He investigated the SU(2) Yang-Mills-Higgs model with an additional Chern-Simon mass term in the action. Pisarski argued that there is a monopole solution that is regular everywhere, but found that it does not possess finite action. There were no exact or numerical solutions being presented by Pisarski. Hence it is our purpose to further investigate this solution in more detail. We obtained numerical regular solutions that smoothly interpolates between the behavior at small and large distances for different values of Chern-Simon term strength and for several fixed values of Higgs field strength. 
\end{abstract}


\section{Introduction}
The study of 2+1 dimemsional gauge theories with a Chern-Simons (CS) term added to the Lagrangian density was first proposed by Deser et al. \cite{kn:1}. They were motivated by the connection of 2+1 dimensional gauge theories to the high temperature limit of four dimensional models. They observed that the massless gauge
field which is spinless becomes massive with spin 1 and the quantum
theory is also infrared and ultraviolet finite in perturbation
theory when interacting with fermions.  Furthermore topology also
leads to a quantization condition on the coupling constant-mass
ratio,  $\frac{4\pi\xi}{g^2}=  n$, where the integer $n = 0, \pm 1,  \pm 2, \ldots$ , in the non-Abelian
gauge theories.

Various non-Abelian gauge theories were studied by various authors in the 1980's. One of these is the work by Pisarski \cite{kn:2}. In his work on the 2+1 dimensional Yang-Mills-Higgs (YMH) theory with the CS term in the Euclidean three space, he noticed that the one-monopole solution analogous to that of the 't Hooft-Polyakov one-monopole solution of the SU(2) Yang-Mills-Higgs (YMH) field theory \cite{kn:3} exist. This famous 't Hooft-Polyakov one-monopole solution with nonzero Higgs mass and self-interaction is a numerically spherically symmetric, finite energy solution in 3+1 dimensions. Contrary to Polyakov \cite{kn:4}, who has shown that in 2+1 dimensions without the CS term, a dilute plasma of monopoles produces confinement of external U(1) charges, Pisarski found that it is the monopoles that are confined and not the external electric charges with the CS included in the theory. He also remark that the analogous monopole solution that he had found, has no counterpart in 3+1 dimensions. Pisarski argued that there is a monopole solution that is regular everywhere, but found that it does not possess finite action. There were no exact or numerical solutions being presented by Pisarski. Hence it is our purpose to further investigate this solution in more detail. We obtained numerical regular solutions that smoothly interpolates between the behavior at small and large distances, for different values of Chern-Simon term strength and for several fixed values of Higgs field strength.

In the next section, we will discuss briefly the 2+1 dimensional YMH theory with the CS term. In section 3, we discuss the magnetic one-monopole ansatz and in section 4, we present our numerical one-monopole solutions. We end with some remarks in the last section.


\section{The SU(2) Yang-Mills-Higgs Theory with the Chern-Simons term}
The SU(2) YMH Lagrangian with non vanishing  Higgs potential is given by
\begin{equation}
{\cal L}_{YMH} = -\frac{1}{4}F^a_{\mu\nu} F^{a\mu\nu} + \frac{1}{2}D^\mu \Phi^a D_\mu \Phi^a - \frac{1}{4}\lambda(\Phi^a\Phi^a - \frac{\mu^2}{\lambda})^2. 
\label{eq.1}
\end{equation}

\noindent Here the Higgs field mass is $\mu$ and the strength of the Higgs potential is $\lambda$. The vacuum expectation value of the Higgs field is $\zeta=\mu/\sqrt{\lambda}$. The YMH Lagrangian (\ref{eq.1}) is gauge invariant under the set of independent local SU(2) transformations at each space-time point.
The covariant derivative of the Higgs field and the gauge field strength tensor are given respectively by 
\begin{eqnarray}
D_{\mu}\Phi^{a} &=& \partial_{\mu} \Phi^{a} + g\epsilon^{abc} A^{b}_{\mu}\Phi^{c},\nonumber\\
F^a_{\mu\nu} &=& \partial_{\mu}A^a_\nu - \partial_{\nu}A^a_\mu + g\epsilon^{abc}A^b_{\mu}A^c_\nu.
\label{eq.2}
\end{eqnarray}

In the non-Abelian SU(2) YMH theory, the inclusion of the CS term leads to the Lagrangian
density which is given by
\begin{equation}
{\cal L} = {\cal L}_{YMH} + {\cal L}_{CS}
\label{eq.3}
\end{equation}
where the YMH Lagrangian density is given as in Eq.(\ref{eq.2})
and the CS Lagrangiann density is given by 
\begin{eqnarray}
{\cal L}_{CS} = \frac{1}{2}\xi\epsilon^{\mu\nu\alpha}
\left(A_\mu^a\partial_\nu A_\alpha^a+\frac{1}{3}A_\mu^a A_\nu^b A_\alpha^c\right).
\label{eq.4}
\end{eqnarray}
The CS parameter $\xi$  and the square of
the gauge couping constant $g$ have the dimension of mass. The CS parameter $\xi$ is real
in Minkowski space and is replaced by $-i\xi$ in the Euclidean space. The SU(2) group indices $a$, $b$, $c$ run from 1 to 3 and the space-time
indices  $\mu$, $\nu$, $\alpha$  = 1, 2 and 0 in the Minkowski space
and 1,  2 and 3 in the Euclidean space.  The
Minkowski space metric used here is $g_{\mu\nu}   = (+ + -)$.

The equations of motion that result from the above Lagrangian density Eq.(\ref{eq.3}) are gauge covariant,
\begin{eqnarray}
D^{\mu}F^a_{\mu\nu} + \frac{1}{2}\xi\epsilon_{\nu\mu\alpha}F^{\mu\alpha a} = \epsilon^{abc}\Phi^{b}D_{\nu}\Phi^c,\nonumber\\
\mbox{where} ~~D^{\mu}F^a_{\mu\nu}=\partial^{\mu}F^a_{\mu\nu} + g\epsilon^{abc}A^{b\mu}F^c_{\mu\nu},
\label{eq.5}\\
\mbox{and}~~ D^{\mu}D_{\mu}\Phi^a = -\lambda\Phi^a(\Phi^{b}\Phi^{b} - \frac{\mu^2}{\lambda}),
\label{eq.6}
\end{eqnarray}
even though the Lagrangian density (\ref{eq.3}) is not gauge invariant.  Under a finite gauge transformation $\omega(x)$, where the gauge potentials are changed by,
\begin{equation}
A_\mu \rightarrow A_\mu' = \omega A_\mu \omega^{-1} -
(\partial_\mu\omega) \omega^{-1},
\label{eq.7}
\end{equation}
the action $S = \int d^3 x {\cal L}$ transforms as  
\begin{eqnarray}
S \rightarrow S' & = & S + \frac{\xi}{g^2}\int d^3
  x ~\epsilon^{\alpha\mu\nu}
  \mbox{Tr} \partial_\mu(A_\alpha\omega^{-1}\partial_\nu\omega)\nonumber\\
& & \mbox{}+ \frac{\xi}{3g^2}\int d^3x ~\epsilon^{\alpha\beta\gamma}\mbox{Tr} (\omega^{-1}
\partial_\alpha\omega\omega^{-1}\partial_\beta\omega\omega^{-1}\partial_\gamma
\omega).
\label{eq.8}
\end{eqnarray}
In order for the first surface integral term in Eq.(\ref{eq.8}) to vanish,
$\omega(x)$ must tend to the identity at time and spatial infinity, that is $\omega(x) \rightarrow I$.
By explicitly writing $\omega(x)$ to be \cite{kn:1}
\begin{eqnarray}
\omega(x) & = & \exp\left(-\theta^a(x)T^a\right),
\label{eq.9}\\
S' & \rightarrow & S + \frac{8\pi^2\xi}{g^2}W(\omega), \nonumber\\
W(\omega) & = & \frac{1}{24\pi^2}\int d^3x ~\epsilon^{\alpha\beta\gamma}
\mbox{Tr} \left(\omega^{-1}\partial_\alpha\omega\omega^{-1}\partial_\beta
\omega\omega^{-1}\partial_\gamma\omega\right)\nonumber\\
& = & \frac{-1}{32\pi^2}\int d^3 x ~\epsilon^{\alpha\beta\gamma}
\epsilon^{abc}\partial_\alpha\left\{\theta^a\partial_\beta
\theta^b\partial_\gamma\theta^c\frac{1}{\theta^2}\left(1 +
\frac{sin\theta}{\theta}\right)\right\},
\label{eq.10}
\end{eqnarray}
where $\theta^2=\theta^a\theta^a$ and $W(\omega)$
which is an integer is the winding
number of the gauge transformation $\omega$. Here $T^a=\frac{\sigma^a}{2i}$ is the SU(2) gauge transformation generators and $\sigma^a$ are the Pauli matrices. The winding number $W(\omega)$ characterizes the
homotopic equivalent class of $\omega(x)$ and vanishes only when
$\omega(x)$
can be continuously deformed to $I$.  The quantization of $W(\omega)$ is
valid both in the Euclidean as well as in the Minkowski space
because it is a metric-independent coordinate invariant.

Hence
the action $S$ is not gauge invariant but the exponential of the
action, $\exp\left(i\int d^3 x {\cal L} \right)$ has to be gauge invariant. 
Therefore $8\pi^2\xi/g^2$  must be equal to $2\pi n$ giving
\begin{equation}
4\pi\xi/ g^2  = n, ~~~n = 0,\pm 1,\pm 2 \ldots
\label{eq.11}
\end{equation}
The same is true in the Euclidean space as the requirement that $\exp\int  d^3 x {\cal L}$  be gauge
invariant still holds as $\xi\rightarrow -i\xi$ in the Euclidean space.     

The dual field which is defined by
\begin{equation}
{}^\ast\!F^{\mu a} = \frac{1}{2}\epsilon^{\mu\alpha\beta} F_{\alpha\beta}^a
\label{eq.12}
\end{equation}
satisfies the Bianchi identity $D^\mu {}^\ast\!F_\mu^a = 0$.
The usual physical quantities which
characterise the solutions are given as follows:
\begin{eqnarray}
\mbox{Magnetic field:} & B^a & = -\frac{1}{2}\epsilon_{ij}F^{aij}
\label{eq.13}
\\
\mbox{Electric field:} & E_i^a & = F_{0i}^a
\label{eq.14}
\\
\mbox{Energy-momentum tensor:} &\theta^{\mu\nu} &= F^{\mu\alpha a}
F_{\alpha a}^\nu  + g^{\mu\nu} L_{YM}
\label{eq.15}
\\
\mbox{Angular momentum:} & J &= \int d^2 x \epsilon_{ij}x^i\theta^{0j}
\label{eq.16}
\\
\mbox{Magnetic flux:} & \Phi^a &= \eta^a\int d^2x B^b\eta^b
\label{eq.17}
\\
\mbox{Electric charge:} & Q^a & = \eta^a\int d^2x(\partial^iE_i^b)\eta^b
\label{eq.18}
\end{eqnarray}   
where $i, j = 1, 2$ and $\eta^a(x)$ is
a unit vector in the internal group space that transform gauge
covariantly.  In all our discussion we set the gauge coupling
constant $g$ to unity.


\section{The Magnetic One-Monopole Ansatz}

The magnetic one-monopole ansatz used in Euclidean space is \cite{kn:2}
\begin{eqnarray}
g A^a_\mu &=& \frac{1}{r}(1-\tau_1(r))(\hat{\theta}^a\hat{\phi}_\mu - \hat{\phi}^a\hat{\theta}_\mu) + \frac{1}{r}\tau_2(r)(\hat{\theta}^a\hat{\theta}_\mu + \hat{\phi}^a\hat{\phi}^\mu) + \hat{r}^a\hat{r}_\mu A(r), ~~~\nonumber\\
g \Phi^a &=& \Phi(r) \hat{r}^a.
\label{eq.19}
\end{eqnarray}
where ~$r=\sqrt{x_1^2+x_2^2+x_3^2}$~ and the spatial spherical coordinate orthonormal unit vectors are defined by
\begin{eqnarray}
\hat{r}_\mu &=& \sin\theta ~\cos \phi ~\delta_{\mu1} + \sin\theta ~\sin \phi ~\delta_{\mu2} + \cos\theta~\delta_{\mu3}, \nonumber\\
\hat{\theta}_\mu &=& \cos\theta ~\cos \phi ~\delta_{\mu1} + \cos\theta ~\sin \phi ~\delta_{\mu2} - \sin\theta ~\delta_{\mu3}, \nonumber\\
\hat{\phi}_\mu &=& -\sin \phi ~\delta_{\mu1} + \cos \phi ~\delta_{\mu2},
\label{eq.20}
\end{eqnarray}
The ansatz (\ref{eq.19}) is the most general ansatz that is invariant under combined rotations of isospin and space-time and under the gauge transformation 
\begin{eqnarray}
\omega(x) = \exp\left(-\theta^a(x)T^a\right), ~~~\theta^a=f(r)\hat{r}^a,
\label{eq.21}
\end{eqnarray}
the gauge potentials transformed as in Eq.(\ref{eq.7}) giving
\begin{eqnarray} 
\tilde{\tau}_1 &=& \tau_1\cos f + \tau_2 \sin f, \nonumber\\
\tilde{\tau}_2 &=& -\tau_1\sin f + \tau_2 \cos f, \nonumber\\
\tilde{A} &=& A - \frac{df}{dr}.
\label{eq.22}
\end{eqnarray}
The total action $S$ in the three dimensional Euclidean space is given by
\begin{eqnarray} 
S&=& S_{YMH} + S_{CS}, \nonumber\\
S_{YMH} &=& \int d^3 x \left\{-\frac{1}{4}F^a_{\mu\nu} F^{a\mu\nu} + \frac{1}{2}D^\mu \Phi^a D_\mu \Phi^a - \frac{1}{4}\lambda(\Phi^a\Phi^a - \frac{\mu^2}{\lambda})^2\right\}\nonumber\\
&=& \frac{4\pi}{g^2}\int_0^\infty dr \left\{(\tau_1^\prime + A\tau_2)^2 + (\tau_2^\prime - A\tau_1)^2 + \frac{1}{2r^2}(1-\tau_1^2-\tau_2^2)^2\right\} \nonumber\\
&+& \frac{4\pi}{g^2}\int_0^\infty dr \left\{\frac{r^2}{2}\Phi^{\prime 2} + \Phi^2(\tau_1^2 + \tau_2^2) - \frac{\mu^2}{2}r^2\Phi^2 + \frac{\lambda}{4}r^2\Phi^4\right\},\nonumber\\
S_{CS} &=& \int d^3x \left\{-\frac{i}{2}\xi\epsilon^{\mu\nu\alpha}
\left(A_\mu^a\partial_\nu A_\alpha^a+\frac{1}{3}A_\mu^a A_\nu^b A_\alpha^c\right)\right\} \nonumber\\
&=& \frac{4\pi}{g^2}\int_0^\infty dr ~i\xi \left\{\tau_1^\prime\tau_2 + \tau_2^\prime(1-\tau_1)- A(1-\tau_1^2-\tau_2^2)\right\}.
\label{eq.23}
\end{eqnarray}
Under the gauge transformation of Eq.(\ref{eq.22}), the action changes by
\begin{eqnarray}
\tilde{S}_{CS} = S_{CS} + \left.\frac{4\pi\xi}{g^2}i\left\{\tau_2(\cos f - 1) - \tau_1\sin f + f \right\}\right|^\infty_0.
\label{eq.24}
\end{eqnarray}
The gauge transformations for which $f(0)=2\pi n_1$ and $f(\infty)=2\pi n_2$, where $n_1$ and $n_2$ are integers, do not change $\tau_1$ and $\tau_2$ at $r=0$ and $r=\infty$. However the action (\ref{eq.23}) changes by $i\frac{8\pi^2\xi}{g^2}n$, ~$n$=integer. Since $\exp(iS)$ is invariant, we have the quantization condition $\frac{4\pi\xi}{g^2}$=integer, that is Eq.(\ref{eq.11}). Hence the presence of monopoles do not change the quantization condition.

Substituting the ansatz (\ref{eq.19}) into the equations of motion (\ref{eq.5}) and (\ref{eq.6}) and letting $\tau_1=\tau$, and $\tau_2=0$, the equations of motion (\ref{eq.5}) and (\ref{eq.6}) reduced to only two coupled differential equations
\begin{eqnarray}
&&\tau^{\prime\prime}-\frac{1}{r^2}\tau(\tau^2-1)-\frac{\xi^2}{4}\left(\frac{\tau^4-1}{\tau^3}\right)-\Phi^2\tau=0,
\label{eq.25}\\
&&\Phi^{\prime\prime} + \frac{2}{r}\Phi^{\prime} - \frac{2}{r^2}\Phi\tau^2 + \Phi(\mu^2-\lambda\Phi^2)=0,
\label{eq.26}\\
&&\mbox{with} ~~~A=\frac{i\xi}{2}\left(\frac{1-\tau^2}{\tau^2}\right).
\label{eq.27}
\end{eqnarray}
Pisarski \cite{kn:2} claimed that there is a regular U(1) monopole solution of the above equations of motion (\ref{eq.25}) - (\ref{eq.27}). However he did not solved these equations of motion. In this paper, we solved these equations of motion numerically and prove that these monopole solutions do exist. 

These regular solutions are monopole solutions as the nonzero Higgs field points in the $\hat{r}^a$ direction in internal space. This is in accordance to the electromagnetic field as proposed by 't Hooft \cite{kn:3},
\begin{eqnarray}
F_{\mu\nu} &=& \hat{\Phi}^a F^a_{\mu\nu} - \epsilon^{abc}\hat{\Phi}^{a}D_{\mu}\hat{\Phi}^{b}D_{\nu}\hat{\Phi}^c,\nonumber\\
	&=& \partial_{\mu}A_\nu - \partial_{\nu}A_\mu - \epsilon^{abc}\hat{\Phi}^{a}\partial_{\mu}\hat{\Phi}^{b}\partial_{\nu}\hat{\Phi}^c,
\label{eq.28}
\end{eqnarray}
\noindent where $A_\mu = \hat{\Phi}^{a}A^a_\mu$, the Higgs unit vector, $\hat{\Phi}^a = \Phi^a/|\Phi|$, and the Higgs field magnitude $|\Phi| = \sqrt{\Phi^{a}\Phi^{a}}$. Hence the electromagnetic field here according to Eq.(\ref{eq.28}) is
\begin{eqnarray}
F_{\mu\nu} = (\hat{\theta}_\mu\hat{\phi}_\nu - \hat{\phi}_\mu\hat{\theta}_\nu)\frac{\hat{r^a}}{r^2},
\label{eq.29}
\end{eqnarray}
which is the magnetic field of a point monopole.

The boundary conditions that we consider for our numerical calculations are 
\begin{eqnarray}
&&\tau \rightarrow 1, ~~A\rightarrow 0,~~\Phi \rightarrow r, ~~ \mbox{as}~~ r \rightarrow 0, 
\label{eq.30}\\
&&\tau \rightarrow \tau_\infty=\left\{\frac{\xi^2}{\xi^2 + 4\zeta^2}\right\}^{1/4}, ~~A\rightarrow \frac{i}{2}\left(\sqrt{\xi^2+4\zeta^2}-\xi\right), ~~\nonumber\\
&&\Phi \rightarrow \zeta=\sqrt{\frac{\mu^2}{\lambda}} ~~\mbox{as} ~~r\rightarrow\infty.
\label{eq.31}
\end{eqnarray} 

\begin{figure}[tbh]
	\centering
		\includegraphics[scale=0.65]{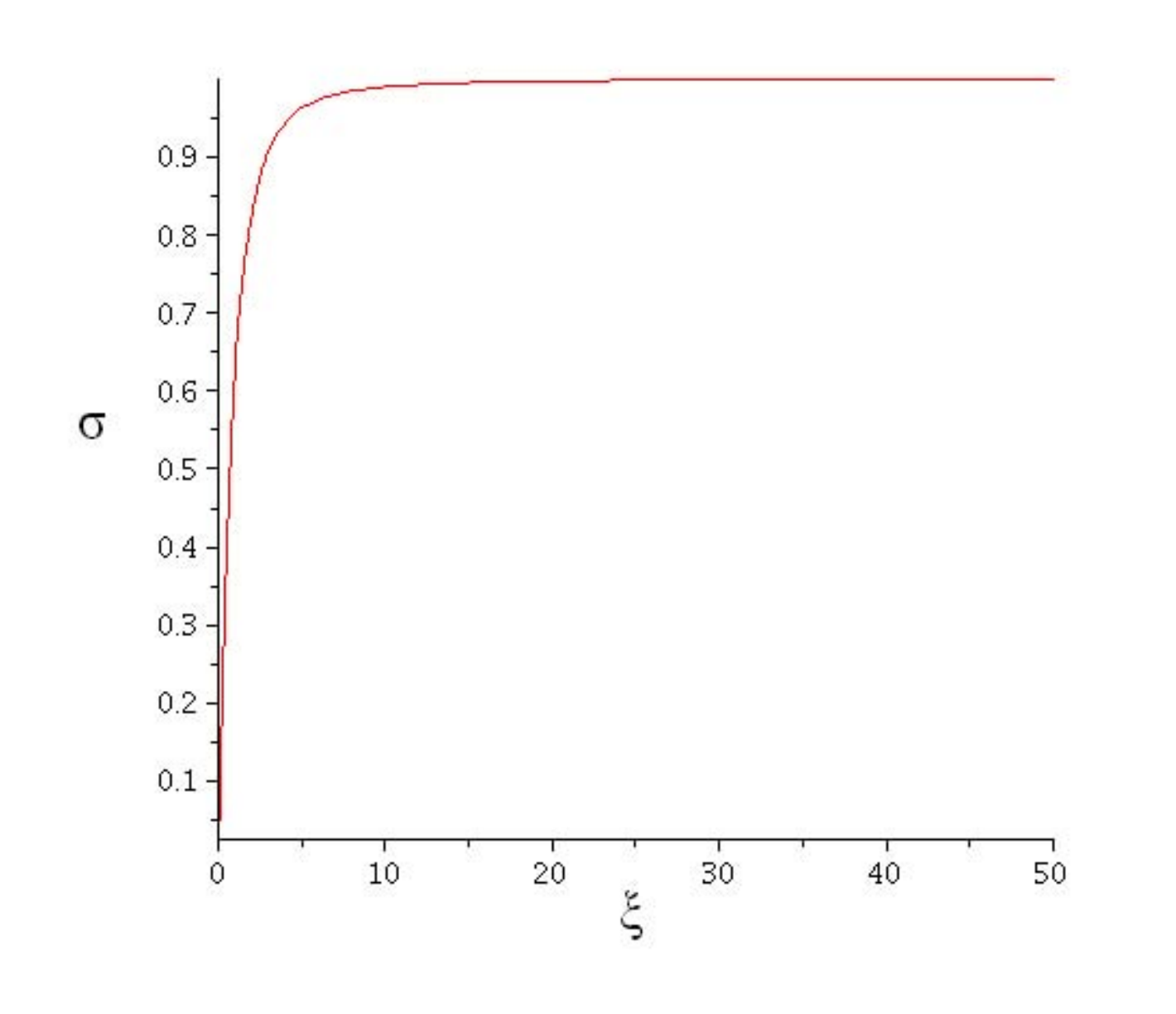}
	\caption{The plot of the string tension $\sigma$ versus the CS parameter $\xi$ in units of $\frac{4\pi}{g^2}$ assuming that $\xi$ can be plotted continuously.}
	\label{fig.1}
\end{figure}

Pisarski cannot find an exact solution with the above boundary conditions (\ref{eq.30}) and (\ref{eq.31}). However, he did mentioned that surely there is a regular solution that smoothly interpolates between the behavior found at large and small $r$  \cite{kn:2}. He also mentioned that while this solution is regular, it does not possess finite action as the total action,
\begin{eqnarray}
S&=& \frac{4\pi}{g^2}\int^R_0 \left\{\tau^{\prime 2}+\frac{(1-\tau^2)^2}{4}\left(\frac{2}{r^2}+\frac{\xi^2}{\tau^2}\right)+\frac{r^2}{2}\Phi^{\prime 2} + \Phi^2\tau^2 - \frac{\mu^2}{2}r^2\Phi^2 + \frac{\lambda}{4}r^2\Phi^4\right\}dr\nonumber\\
&=& \frac{4\pi}{g^2}\int^R_0 \left\{\Phi^2\tau^2+\frac{\xi^2}{4}\left(\frac{1-\tau^2}{\tau}\right)^2+.....\right\}dr,
\label{eq.32}
\end{eqnarray}
is proportional to $\sigma R$ where is the radius of space-time and $\sigma$ is the string tension for a monopole-antimonopole pair and is given by
\begin{eqnarray}
\sigma = \frac{4\pi}{g^2} \left\{\zeta^2\tau_\infty^2+\frac{\xi^2}{4}\left(\frac{1-\tau_\infty^2}{\tau_\infty}\right)^2\right\}.
\label{eq.33}
\end{eqnarray}
The plot in Figure 1 shows how the string tension $\sigma$ varies with the CS parameter $\xi$ in units of $\frac{4\pi}{g^2}$ when $\zeta=1$. 
Hence for small CS parameter $\xi\ll \zeta$, ~~$\sigma \approx \frac{4\pi}{g^2}\xi\zeta$~~ and it increases proportionally with $\xi$. However for large CS parameter $\xi\gg \zeta$, ~~$\sigma \rightarrow \frac{4\pi}{g^2}\zeta^2$. Therefore the presence of the CS term not only causes all gauge fields, including the photons, to be massive but also causes monopole and antimonopole to be tightly bound together. However as argued by d'Hoker and Vinet \cite{kn:5}, since there is no long range correlations, topologically massive theories do not confine external charges but instead monopoles are confined. 

\section{The Numerical One-Monopole Solutions}

In this paper the numerical calculations were performed using the Maple 12 and MatLab 2007 softwares. The two second order equations of motion (\ref{eq.5}) - (\ref{eq.6}) are reduced to two differential equations (\ref{eq.25}) - (\ref{eq.26}) with the ansatz (\ref{eq.19}). These two differential equations are then transformed into a system of nonlinear equations by using the finite difference approximation. They are then discretized into 50 divisions in the $\bar{x}$ coordinate where $\bar{x}=\frac{r}{r+1}$ is the so-called finite interval compactified coordinate. By considering the $\bar{x}$ coordinate, the derivative with respect to the radial coordinate is then replaced accordingly by $\frac{d}{dr} \rightarrow (1-\bar{x})^2 \frac{d}{d\bar{x}}$, ~~$\frac{d^2}{dr^2} \rightarrow (1-\bar{x})^4\frac{d^2}{d\bar{x}^2} - 2(1-\bar{x})^3\frac{d}{d\bar{x}}$. The system of nonlinear equations are then solved numerically with the Gauss-Newton method by providing the solver with good initial guess. 

\begin{figure}[tbh]
	\centering
		\includegraphics[scale=0.45]{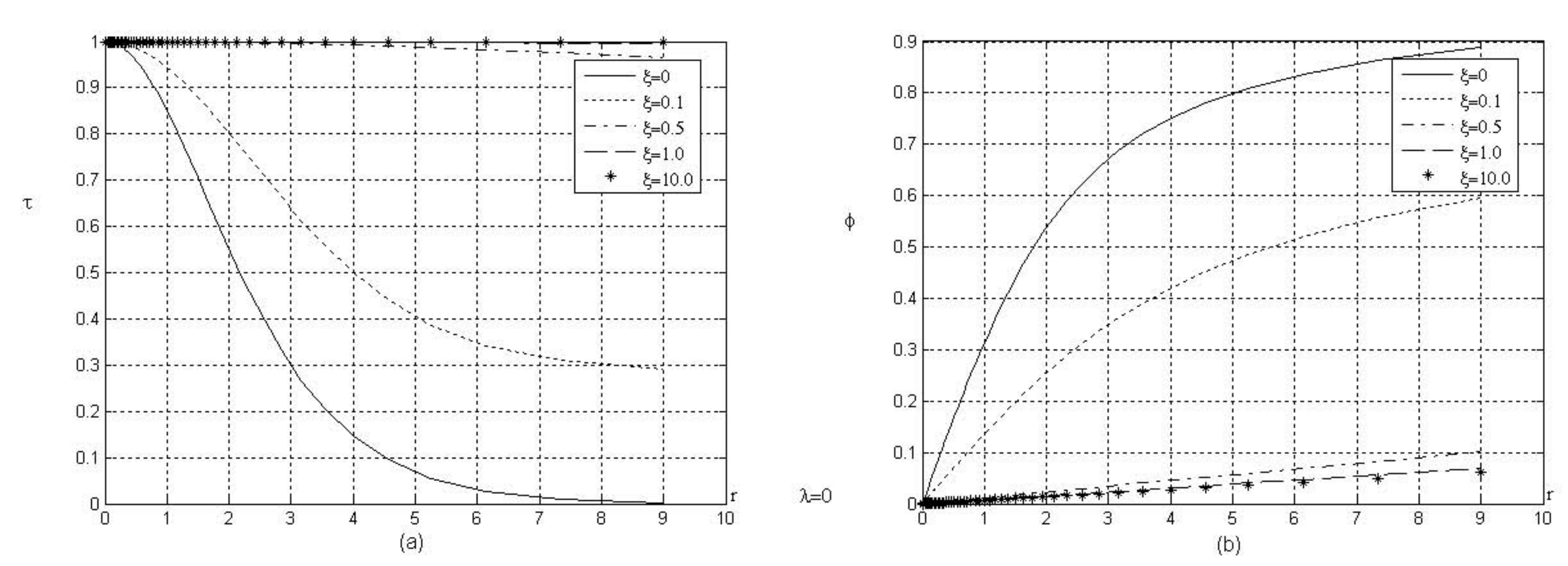}
	\caption{(a) The plot of $\Phi(r)$ versus $r$ and (b) the plot of $\tau(r)$ versus $r$ for different values of CS parameter $\xi= 0, 0.1, 0.5, 1$ and 10, when $\lambda=0$.}
	\label{fig.2}
\end{figure}

\begin{figure}[tbh]
	\centering
		\includegraphics[scale=0.45]{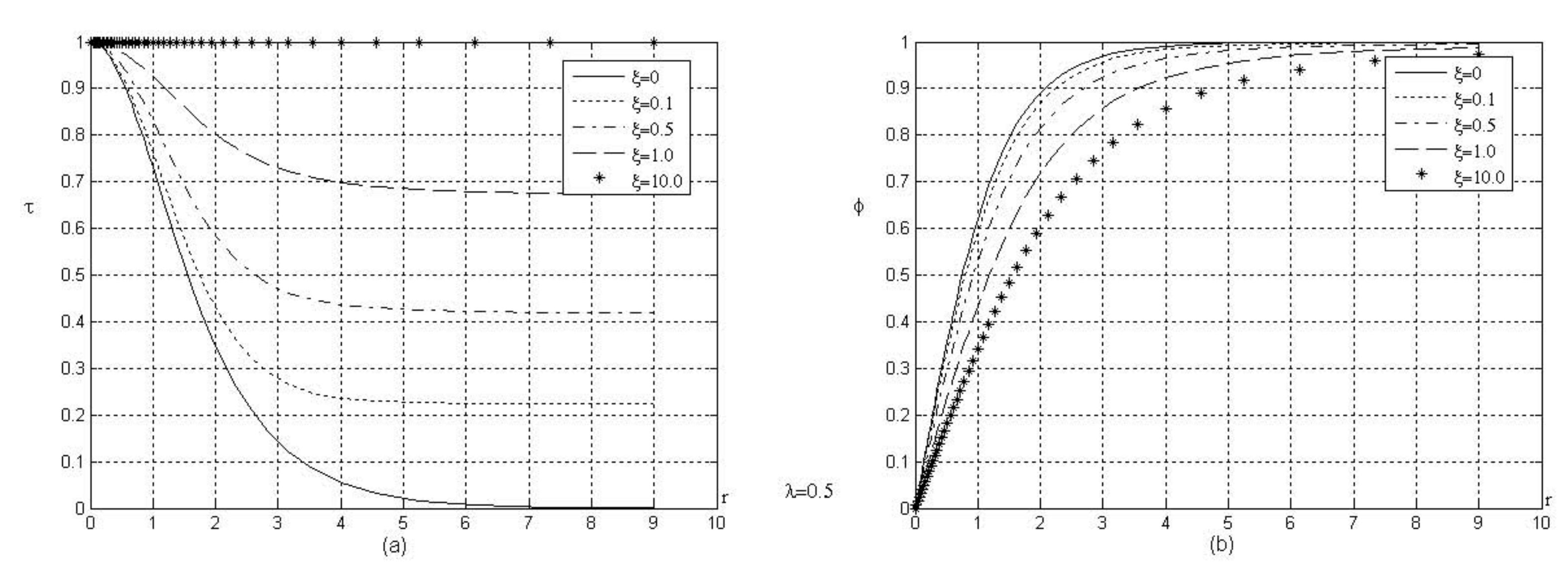}
	\caption{(a) The plot of $\Phi(r)$ versus $r$ and (b) the plot of $\tau(r)$ versus $r$ for different values of CS parameter $\xi= 0, 0.1, 0.5, 1$ and 10, when $\lambda=0.5$.}
	\label{fig.3}
\end{figure}

\begin{figure}[tbh]
	\centering
		\includegraphics[scale=0.45]{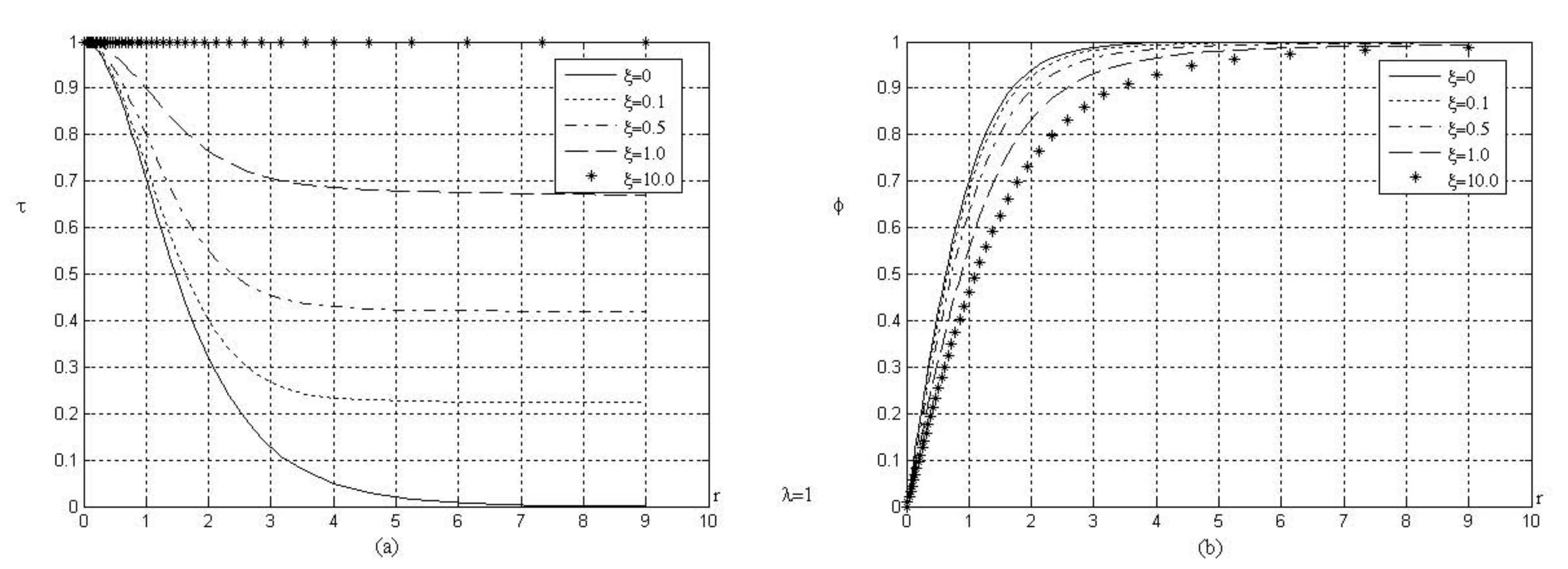}
	\caption{(a) The plot of $\Phi(r)$ versus $r$ and (b) the plot of $\tau(r)$ versus $r$ for different values of CS parameter $\xi= 0, 0.1, 0.5, 1$ and 10, when $\lambda=1$.}
	\label{fig.4}
\end{figure}

\begin{figure}[tbh]
	\centering
	 \vspace{3.0cm}
		\includegraphics[scale=0.45]{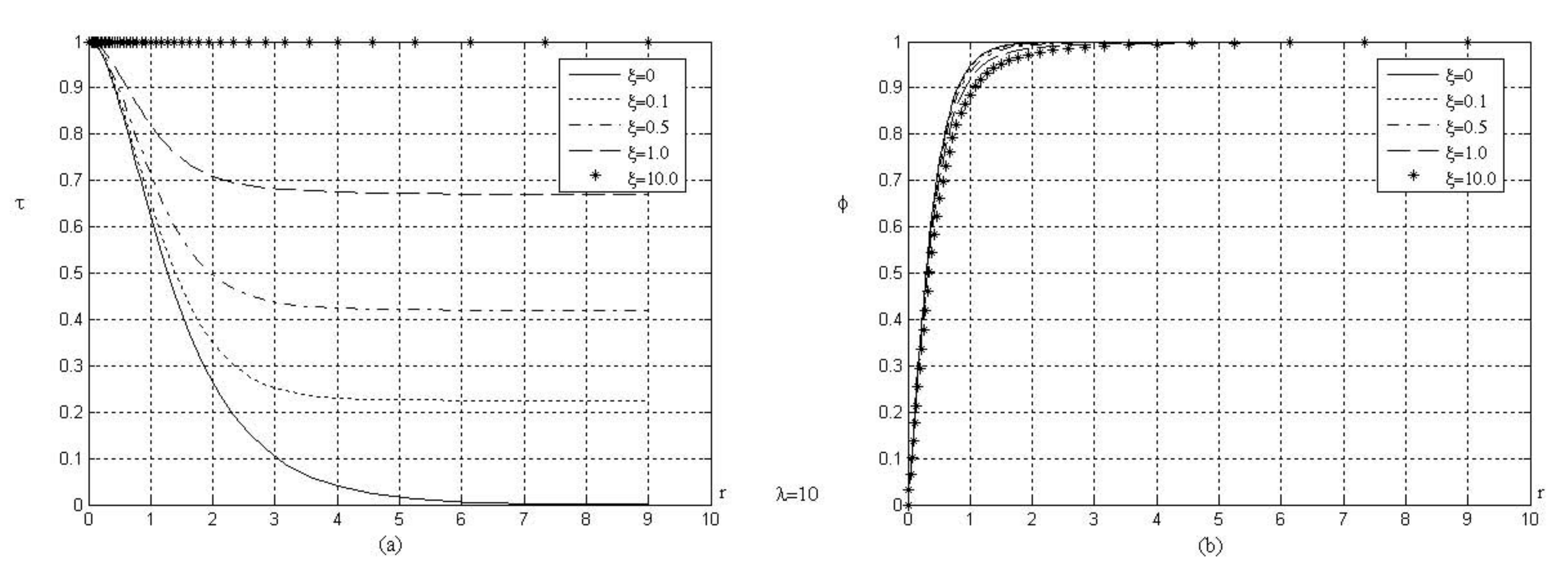}
	\caption{(a) The plot of $\Phi(r)$ versus $r$ and (b) the plot of $\tau(r)$ versus $r$ for different values of CS parameter $\xi= 0, 0.1, 0.5, 1$ and 10, when $\lambda=10$.}
	\label{fig.5}
\end{figure}

When the expectation value of the Higgs field is $\zeta=1$, and $\lambda = 0, \frac{1}{2}, 1$, and 10, we solved the two differential (\ref{eq.25}) - (\ref{eq.26}) numerically for $\tau$ and $\Phi$, for different values of CS parameter $\xi = 0, 0.1, 0.5, 1$ and 10. These numerical solutions are shown graphically in Figure (\ref{fig.2}) - (\ref{fig.5}). We can conclude from the graphs of the numerical solutions that with increasing values of the strength of the Higgs potential $\lambda$, the Higgs field approaches its expectation value $\zeta=1$ faster. However the effect of the CS parameter slows down the rate of $\Phi$ approaching $\zeta$. Similarly with increasing values of the strength of the Higgs potential $\lambda$, the function $\tau \rightarrow \tau_\infty$ faster. The asymptotic value of $\tau$ at large $r$, that is $\tau_\infty$, however now varies with the CS parameter $\xi$ as $\tau_\infty=\left\{\frac{\xi^2}{\xi^2 + 4\zeta^2}\right\}^{1/4}$.


\section{Remarks}
As mentioned by Pisarski \cite{kn:2} there are four solutions to Eq.(\ref{eq.25}) at large $r$, as ~~$\tau \rightarrow \pm\tau_\infty, ~\pm i \tau_\infty$. However the imaginary solutions can give negative values of the string tension $\sigma$ and hence are not physical. Also purely imaginary gauge potentials $A^a_\mu$ will also give negative action density and may not represent a physical vacuum state.

We have found the regular numerical one-monopole solutions of the massive topological gauge theory whose existence was predicted by Pisarski \cite{kn:2} and he interpreted the string tension $\sigma$ as the tension between a monopole-antimonopole pair. However until now there were no monopole-antimonopole pair (MAP) solutions of the 2+1 dimensional YMH gauge field theory with the CS term. Hence our next task is to search for the MAP solutions in this theory. Similar to the MAP solution in 3+1 dimensional YMH gauge field theory, these MAP solution cannot posses spherical symmetry but only axial symmetry.\vspace{3.5cm}

We would also like to note that in the 2+1 dimensional YMH field theory with the CS term, the one-monopole solution presented here is not a solition but instead it is an instanton and may represent some tunnelling events. 

\section*{Acknowledgements}
The authors would like to thank the Ministry of Higher Education Malaysia (Putrajaya) of Malaysia for the award of Fundamental Research Grant Scheme (FRGS) (Account Number: 203/PFIZIK/671168).

\newpage

\end{document}